\documentclass{article}

\usepackage{arxiv}

\usepackage[utf8]{inputenc} % allow utf-8 input
\usepackage[T1]{fontenc}    % use 8-bit T1 fonts
\usepackage{hyperref}       % hyperlinks
\usepackage{url}            % simple URL typesetting
\usepackage{booktabs}       % professional-quality tables
\usepackage{amsfonts}       % blackboard math symbols
\usepackage{nicefrac}       % compact symbols for 1/2, etc.
\usepackage{microtype}      % microtypography

\usepackage[product-units = power]{siunitx}
\DeclareSIUnit\clight{\text{\ensuremath{c}}}
\usepackage[
    backend=biber,
    style=numeric,
    sorting=none
  ]{biblatex}
\usepackage{graphicx}
\graphicspath{{figures/}}

\title{Muon absorption tomography of a lead structure through the use of iterative algorithms}

\author{
  Guglielmo Baccani \\
  Department of Physics and Astronomy\\
  University of Florence\\
  P.zza S.Marco 4, 50121 Florence, Italy \\
  \texttt{guglielmo.baccani@unifi.it} \\
  National Institute for Nuclear Physics (INFN)\\
  Unit of Florence\\
  Via Sansone 1, 50019 Sesto Fiorentino (FI), Italy \\
}

\begin{document}

\maketitle

\begin{abstract}
The muon radiography technique is part of the straight-ray scanning imaging methods and it is usually employed to produce two-dimensional images of the integrated density in the radial direction from the detector position. However, in the literature there are already some examples of muon tomographies made by combining several measurements from different observation points. The work presented in this article fits into this line of research: a tomography of a target composed of lead blocks arranged in a non-trivial geometry was carried out by means of muon absorption radiography measurements. To solve the inversion problem, an iterative algorithm called SART was employed, obtaining better results than using a Chi-squared minimization.
\end{abstract}

% keywords can be removed

\section{Muon radiography and scattering tomography}
Muon radiography (or muography) is a geophysical technique that allows to study the density and the atomic number of large dimension systems taking advantage of cosmic ray muons.

The muon is an unstable lepton with a mean life of \SI{2.2}{\us} and a mass 200 times greater than that of the electron ($m_\mu \approx \SI[per-mode=symbol]{106}{\MeV\per\clight\squared}$). Muons are the most frequent charged particles from cosmic rays at the sea level, with a vertical flux of about \SI{70}{m^{-2}.s^{-1}\steradian^{-1}} for energies greater than \SI{1}{\GeV}~\cite[chapter~30.3]{PhysRevD}. The integral muon flux depends on the zenith angle $\theta$ approximately as $\cos^{2}\theta$, while it weakly depends on the azimuth angle $\phi$ because of the interaction of the mainly positive particles in the primarily cosmic radiation with the Earth's magnetic field (this is the so called East-West asymmetry). Going into more detail, the muons spectrum depends on the measuring site (on its altitude, latitude and longitude) and varies with time: solar activity and variations in the atmospheric pressure influence the flux of cosmic muons~\cite{grieder2001cosmic, bonechi2020atmospheric}.

There are mainly two muographic techniques that are based on two different interaction phenomena of muons within matter: muon transmission (and absorption) radiography and muon scattering tomography. All of these techniques relies on the detection of muon tracks by means of particle detectors called trackers. Some general reviews can be found in~\cite{procureur2018muon, bonechi2020atmospheric}.

\begin{figure}
    \centering
    \includegraphics[width=0.8\linewidth,trim=0 0 0 40,clip]{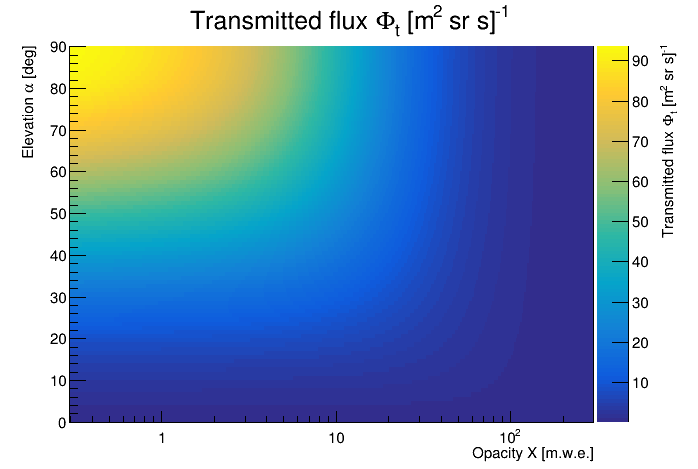}
    \caption{Transmitted muon flux as a function of the crossed opacity (in meter water equivalent) and of the elevation angle (the dependency from the zenith angle is neglected).}
    \label{fig:simulazione_conversione}
\end{figure}

Muon transmission and absorption radiography are based on the energy loss (and the eventual decay) of muons when they interact with matter~\cite[chapter~34.2]{PhysRevD}. A muon with a given impulse $p_\text{min}$, before being stopped, will be able to cross on average a certain opacity $X$ defined as ${X = \int \rho \,dL }$, where $\rho$ is the density of the material and $L$ is the particle path length.
%As a first approximation the opacity of a target  doesn't depend on the material composition.
The relationship between minimum impulse and crossed opacity can be found tabulated in reference~\cite{groom2001muon}. The higher the opacity of the target, the lower the number (and flux) of muons that have an impulse high enough to cross it. Fixed a certain angular direction $(\theta ,\phi)$ and a certain opacity, the expected value of the flux transmitted through the target can be estimated as the integral from $p_\text{min}$ to infinity of the differential flux in that direction. In figure~\ref{fig:simulazione_conversione} the transmitted muon flux is depicted as a function of the opacity of the crossed object and of the elevation angle (the dependence of the muon flux from the azimuth angle is neglected).
This histogram has been obtained from the knowledge of the differential muon flux on Earth measured with the ADAMO spectrometer~\cite{bonechi2005development}. With the previous conversion map or using appropriate calibration measurements (as in this work), the fraction of transmitted (absorbed) muons can be converted into an opacity measurement for each direction of observation. Dividing the opacity by the total target length, one eventually obtains the average density along the inspected direction.

For the transmission technique a single tracker is placed downstream the target to detect the transmitted muons, while for the absorption one the object under inspection is placed between the tracker and a veto plane (respectively above and below the target) to identify those muons that have been stopped within the target. Although the two techniques are similar, transmission muon radiography is the most common as it allows to examine very large objects.

Transmission muography sees its greatest development in the field of volcanology~\cite{tanaka2007high, ambrosino2015joint, tanaka2014radiographic, jourde2016muon, ambrosino2014mu, d2020muon, tioukov2019first, carloganu2013towards}. The first studies were able to provide a description of the internal architecture of volcanoes and afterwards dynamic studies have been carried out, demonstrating the possibility of monitoring volcanoes in real time so as to prevent the risks associated with an eruptive event. Apart from volcanology there are several other applications in the fields of archaeology, geology, mining and civil engineering. Following in the footsteps of Luis Alvarez's pioneering measure inside the Chefren pyramid from the 1970's~\cite{alvarez1970search}, more recently a hidden chamber inside the Cheops pyramid was discovered by the ScanPyramids collaboration~\cite{morishima2017discovery}. In the field of geology the technique has been employed to determine the shape of the bedrock underneath alpine glaciers in Switzerland~\cite{nishiyama2017first}, while there are numerous uses in the mining sector~\cite{baccani2019muon, schouten2018muon, guardincerri20173d, bryman2014muon, lesparre20173}. For all of these applications a typical acquisition lasts from a few days up to a month.

The scattering muon tomography technique takes advantage of another phenomena: the deflection of muons from their straight trajectory due to multiple coulomb interactions~\cite[chapter~34.3]{PhysRevD}. The deflection angle has null average and depends on velocity and impulse of the particle, and on opacity and atomic number $Z$ of the crossed material. The deflection of muons is particularly relevant for low impulse muons and it can constitute a not negligible background for the transmission (absorption) radiography~\cite{bonechi2020atmospheric}. Scattering tomography allows to directly reconstruct the density and the atomic number of the target by measuring the deflection of muons trough it. To do that two tracking detectors, one upstream and one downstream the target, are needed, thus limiting the size of the objects that can be studied (that is limited also by the fact that the technique is no longer sensible when multiple deflections occur). Given the higher level of information obtained from these measurements, acquisition times are generally reduced (around a few minutes) compared to transmission radiography. The technique is particularly suitable to discriminate materials with an high atomic number~\cite{borozdin2003radiographic} opening applications in the nuclear sector and in homeland security~\cite{riggi2018muon, checchia2019infn, poulson2017cosmic, pesente2009first, clarkson2015characterising, ambrosino2015assessing, borozdin2012cosmic, miyadera2013imaging}.

\section{3D reconstruction with muon radiography}

\subsection{Tomographic reconstruction}

While muon scattering tomography is naturally a three-dimensional imaging technique, muon radiography provides bidimensional maps of density integrated on the radial direction from the point of view of the detector. Nonetheless, just like in medical CT, combining multiple muon radiography measures from different positions, it is possible to obtain the tree-dimensional distribution of density.
Some tomographic reconstructions using muon radiography have already been carried out and each has its own peculiarity which depends on the system under observation~\cite{erlandson2016one, lesparre20173, bryman2014muon, rosas2017three, davis2012joint, guardincerri20173d, jourde2015improvement, schouten2018muon}: unlike the work presented in this article, some of these reconstructions involve a combination of muographic and gravimetrical measurements~\cite{davis2012joint, jourde2015improvement, schouten2018muon} and many of them use regularization terms based on a priori information to reduce the non-uniqueness typical of these problems~\cite{bryman2014muon, rosas2017three, davis2012joint, guardincerri20173d, schouten2018muon}.

\subsection{Inversion problem definition}
In all these reconstructions the inspected volume, from now on World, is modeled as a 3D grid of Voxels usually cuboids in shape (except in~\cite{lesparre20173}) with unknown uniform density. Then the muographic information is organized in a series of Rays each characterized by an opacity measurement and by geometrical limits: in case the tracker can be considered point-like, each Ray will correspond to a precise solid angle\label{subsec:ray_solid_angle}.
Supposing to have $I$ Rays and $J$ Voxels, after this discretization process, from the equation of the opacity definition we obtain the following system of equations:
\begin{equation}
    \label{eq:opa_discrete}
    X_i = \sum_{j=1}^J L_{ij} \rho_j =
    \vec L_i \vec \rho\, ,
    \qquad
    i=1, 2, \dots, I,
\end{equation}
where $X_i$ is the opacity of the $i$th Ray, $\rho_j$ is the density of the $j$th Voxel and $L_{ij}$ is the average intersection length  between the $i$th Ray and the $j$th Voxel. For a given Ray, just a few Voxels will be intersected and will contribute with their density to the Ray's opacity and therefore $L_{ij}$ is an element of a sparse matrix ($\mathbf{L}$). The goal will be to determine the density values that solve the system of equations~\ref{eq:opa_discrete}. In case of real opacity measurements with errors $\sigma_i$, one will have to find the density vector that minimizes the Chi-squared
\begin{equation}
    \label{eq:chi2_opa}
    \chi^2 =
    \sum_{i=1}^I \left(
    \frac{X_i - \vec L_i \vec \rho}{\sigma_i} \right) ^2.
\end{equation}

%\subsection{Non uniqueness and regularization methods}
%Do I have to write something about it in your opinion?

%Usually min Chi2 with regularization implying a smoothing constrain.

\subsection{The ART family algorithms}

Given the usually high values for $I$ and $J$ (see section~\ref{subsec:vox_ray_def} for example) the minimization of the Chi-squared defined in~\ref{eq:chi2_opa} is computationally expensive. To reduce the computational time and still achieve a good result, iterative methods can be used. ART (Algebraic Reconstruction Technique) is an iterative algorithm commonly used in medical physics (see~\cite{Civinini_2017} for instance) that allows to solve systems of linear equations even with a large number of unknowns and equations. The vector of unknown densities of dimension $J$ can be represented as a point in a $J$-dimensional space while the system of $I$ equations can be thought as $I$ hyperplanes in the same space (since each equation is expressed as a linear combination of the Voxels densities). In an ideal problem, without any errors, the solution will correspond to the intersection of the hyperplanes. The ART algorithm allows to solve the system by sequentially projecting the point-solution on each hyperplane-equation at each iteration. Expressing this in formula, projecting on the $i$th hyperplane, the vector of the density variation will be given by
\begin{equation}
    \label{eq:ART}
    \Delta \vec \rho_i ^\textup{ART}=
    \frac{X_i - \vec L_i \vec \rho }{|\vec L_i|^2} \, \vec L_i.
\end{equation}
Interesting convergence properties can be found in literature also in the case of real measurements affected by errors~\cite{kak2002principles} and the ART algorithm has been already used to perform a tomographic reconstruction of nuclear reactor using synthetic data~\cite{erlandson2016one}. Nonetheless two aspects are not fully satisfactory: the order chosen for projecting the solution on the hyperplanes can determine the convergence speed and, secondly, there is no way to take account of the error on the opacity measurement of each Ray.

The SART (Simultaneous ART) algorithm is a small variant of the ART algorithm~\cite{kak2002principles}: each single iteration corresponds to the average of the variations that would be obtained by projecting the point on each hyperplane using the ART algorithm. In this way it is possible to disengage from the order of projection and a bigger relevance can be given to the more precise opacity measurements. To do this, a weighted average will be used with weights $w_i$ inversely proportional to the measurement error on opacity. In formulas the density variation vector can be expressed as:
\begin{equation}
    \label{eq:SART}
    \Delta \vec \rho_i ^\textup{SART} = \frac{\sum_{i=1}^{I} w_i \Delta \vec \rho_i ^\textup{ART} }{\sum_{i=1}^{I} w_i},
    \qquad
    w_i = \left|
    \frac{X_i - \vec L_i \vec \rho}{\sigma_i} \right|.
\end{equation}

\section{3D absorption muography test}
The objective of this work is the development and testing of tomographic inversion algorithms for muon radiography measurements. To compare the inversion algorithms presented (SART and Chi-squared minimum) a test measurement was carried out at the Irfu Institute of the CEA centre of Saclay.

\subsection{The detector}
The detector used for this work is the TomoMu tracker: a modular gas-based detector made up with Micromegas (Micromesh Gaseous Structure) tracking planes~\cite{bouteille2016large, bouteille2016micromegas}. Each plane is characterized by: 
\begin{enumerate}
    \item double readout for the X and Y direction;
    \item spatial resolution of about \SI{0.7}{mm} for each view; %\textbf{(do I insert the residual plots?)};
    \item active area of \SI{50 x 50}{\cm}.
\end{enumerate}

%commentare le correzioni sull'allineamento x, y, z???

\begin{figure}
    \centering
    \includegraphics[height=0.27\textheight] {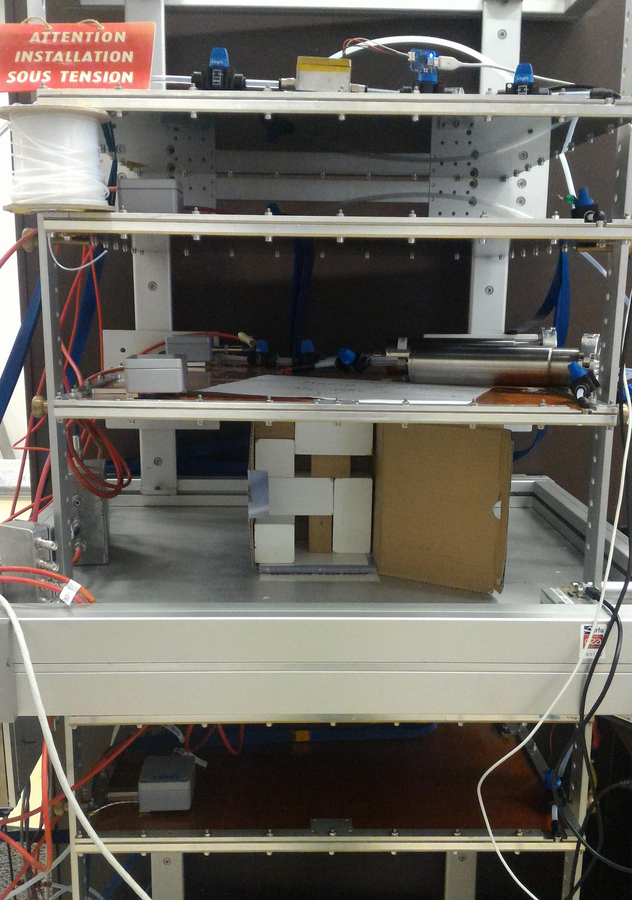}~
    \includegraphics[height=0.27\textheight] {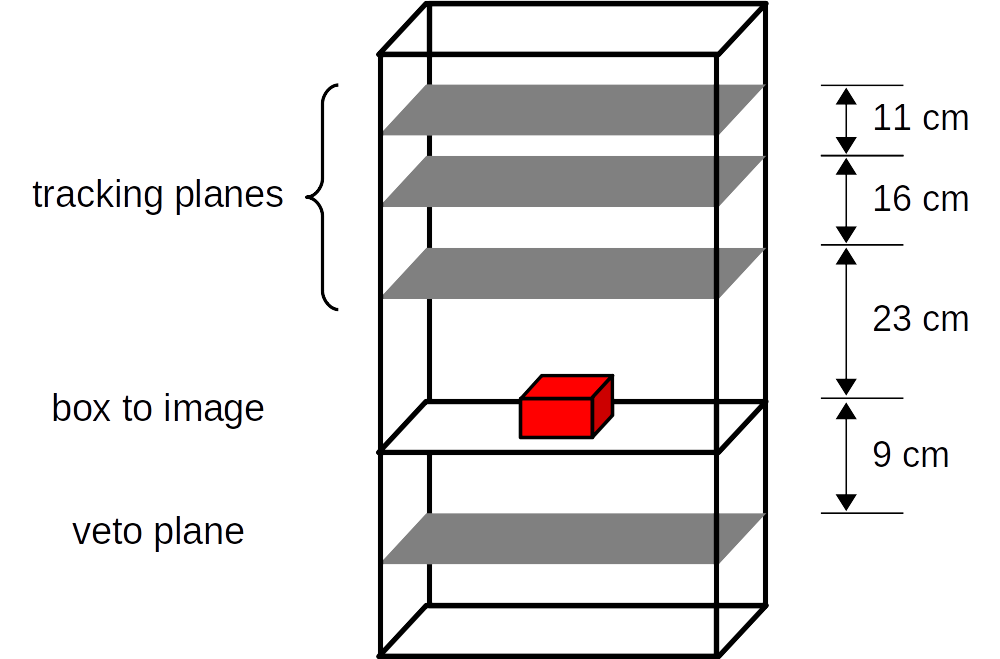}
    \caption{An image of the TomoMu detector and its schematic drawing showing the distances between the planes. From the top to the bottom we see three tracking planes, the aluminum sustaining plane with the target on it and the veto plane.}
    \label{fig:detector}
\end{figure}

For this work we decided to study a small system that could fit inside the detector using absorption muon radiography: the muon tracks were identified by three tracking planes  above the target and a veto plane was positioned under it. In figure~\ref{fig:detector} the TomoMu tracker and its schematic picture are shown side by side.

\subsection{The target measurements}
\begin{figure}
    \centering
    \includegraphics[angle=-90,origin=c, width=0.4\columnwidth]{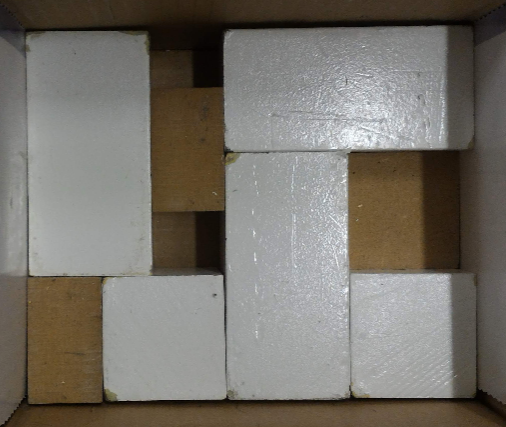}~
    \includegraphics[angle=-90,origin=c, width=0.4\columnwidth]{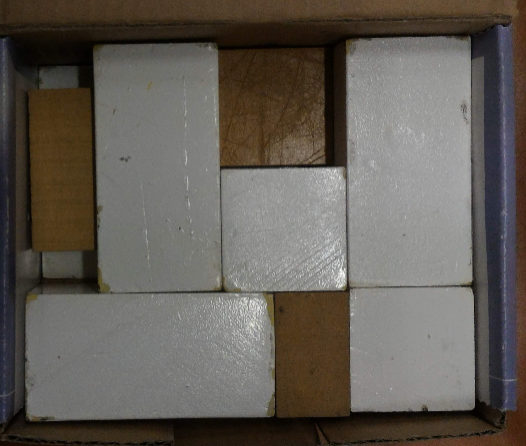}
    \caption{The two layers the target is composed of. The lead bricks (in white) and the wooden wedges (in brown) are visible. From this view both layers are \SI{15}{cm} large, \SI{18}{cm} high and \SI{10}{cm} depth.}
    \label{fig:box_layers}
\end{figure}

Given the detector shape we selected a 3d non-trivial system that could fit on the detector supporting plane as target of the muon tomography: a cardboard box filled with lead bricks. The bricks used for this work (with sizes \SI{5 x 5 x 5}{\cm} or \SI{10 x 5 x 5}{\cm}) were kept separate with some wooden wedges (with negligible density) so that it was possible to rotate the box, keeping the position of the bricks stable. The overall box is \SI{18 x 15 x 10}{\cm} in size and it is composed of two bricks layers as shown in figure~\ref{fig:box_layers}.

With a single measurement the target is already inspected from all the directions, close to the vertical, that fall inside the angular acceptance of the detector. In order to increase the number of measurements, especially in the horizontal direction, we decided to rotate the target in two other orthogonal positions acquiring the corresponding opacity data. This was done to leave the detector stationary, as it would have been rather inconvenient to tilt it, and to avoid being affected by the muon flux decrease at low elevation angles. A total of three different measurements with the target rotated in three orthogonal directions inside the detector were performed. The system remained in acquisition for a total of three weeks (one week for each position).

\subsection{Voxels and Rays geometry definition}

\label{subsec:vox_ray_def}
For this experiment two different reconstruction configurations have been selected. In the first configuration a \SI{19 x 22 x 14}{\cm} size World was chosen, with cubic Voxels with a \SI{1}{\cm} edge (with a total number of Voxels equal to $J=5852$). In the second configuration larger Voxels were chosen, with \SI{3}{\cm} edge, and the World dimension was set to \SI{21 x 24 x 18}{\cm} obtaining 336 Voxels in total ($J=336$). From here on, these two configurations will be referred to as World1 and World3. In both cases the World has been defined in such a way as to completely encompass the target, leaving a few centimeters of air as a margin.

Since the detector cannot be considered point-like compared to the target the Rays cannot be geometrically defined simply as a solid angle (see paragraph~\ref{subsec:ray_solid_angle}). For each Ray a certain number of tracks will be selected, and will be used to define the absorption probability and the opacity measurement. The tracks selection, and consequently the Rays geometry, is obtained defining a series of consecutive intervals (just like for histograms), for the following quantities:
\begin{itemize}
    \item impact coordinates of the tracks on the horizontal median plane of the World $(x_0, y_0)$;
    \item tangent of angle of the partial tracks in the X-Z and Y-Z views $(\tan \alpha_x = \frac{\Delta x}{\Delta z},\, \tan \alpha_y = \frac{\Delta y}{\Delta z})$.
\end{itemize}
For this analysis the following binning dimensions were chosen in order to have a good opacity resolution: $\Delta x_0 = \Delta y_0 = \SI{2}{\cm}$ and $\Delta(\tan \alpha_x) = \Delta(\tan \alpha_y)= 0.2$. With this binning choice, taking advantage of the calibration measurements presented in subsection~\ref{subsec:calibration}, an opacity resolution of \SI{1.3}{\cm} is obtained for a Ray with $\tan \alpha_{x, y} \in [-0.1, 0.1]$ (close to the vertical direction) crossing \SI{3}{\cm} of lead after a week of data taking.

Given the geometrical definition, the total number of Rays that intersect the World was $I=3398$ for the World1 configuration and $I=4095$ for the World3 configuration. The number of Rays in the second configuration is greater since the World itself is larger.

\subsection{Calibration measurements and opacity initialization}
\label{subsec:calibration}
For each configuration and for every Ray, the number of total reconstructed tracks $N_{\text{tot}}$ and the number of absorbed tracks $N_{\text{abs}}$, which were not revealed by the veto plane and had therefore been absorbed (or deflected) in the system under investigation, were recorded.
\begin{figure}
    \centering
    \includegraphics[width=0.8\linewidth,trim=0 0 0 55,clip]{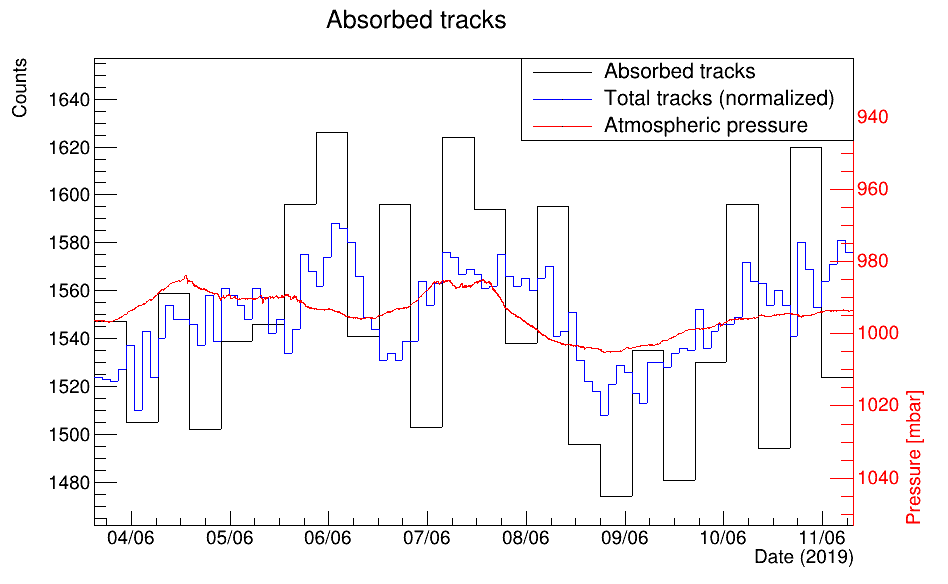}
    \caption{The graph shows the temporal variation of the atmospheric pressure (in red) and of the number of total and absorbed tracks (respectively in blue and black) for a week of data taking. The number of total tracks was normalized by a factor of approximately 20 and a wider binning was chosen for the absorbed tracks to reduce statistical fluctuations. The anti-correlation between the number of reconstructed tracks and the atmospheric pressure is evident (the pressure scale on the right is inverted).}
    \label{fig:muon_rate_pressure}
\end{figure}
As we can see in figure~\ref{fig:muon_rate_pressure}, both $N_{\text{tot}}$ and $N_{\text{abs}}$ seem to be equally dependent on atmospheric pressure. To remove this dependence and reduce any other systematic effects, for each Ray the absorption probability was estimated as $P_{\text{abs}} = N_{\text{abs}} / N_{\text{tot}}$. Since the number of absorbed tracks $N_{\text{abs}}$ can be close to zero, a Bayesian approach is used to calculate the error on the absorption probability~\cite{ullrich2007treatment}.

Three calibration measurements (lasting one day each) were carried out with uniform layers of lead with a thickness of \SIlist[list-units = single]{0;5;10}{\cm}. Compared to the target measurements, a shorter acquisition time was used for the calibration measurements since it was possible to integrate the muon tracks over the entire uniform layer of lead thus obtaining a lower statistical uncertainty. These calibrations allowed to convert the absorption probability of each Ray into an opacity measure assuming a proportionality relationship between the absorbed flux and the crossed opacity (or the lead thickness). This assumption is reasonable since the energy loss of a minimum ionizing muon in \SI{10}{\cm} of lead is of about \SI{100}{\MeV} and, for these energy scale, the differential muon flux is almost flat~\cite{bonechi2005development}.  Consequently the lead thickness will be proportional to the number of absorbed tracks $N_{\text{abs}}$ and also to the absorption probability $P_{\text{abs}}$ (since $P_{\text{abs}} \ll 1$ for the selected target). Increasing the target thickness, the absorption probability would tend asymptotically to one, loosing its initial linear trend.
\begin{figure}
    \centering
    \includegraphics[width=0.8\linewidth,trim=0 0 0 30,clip]{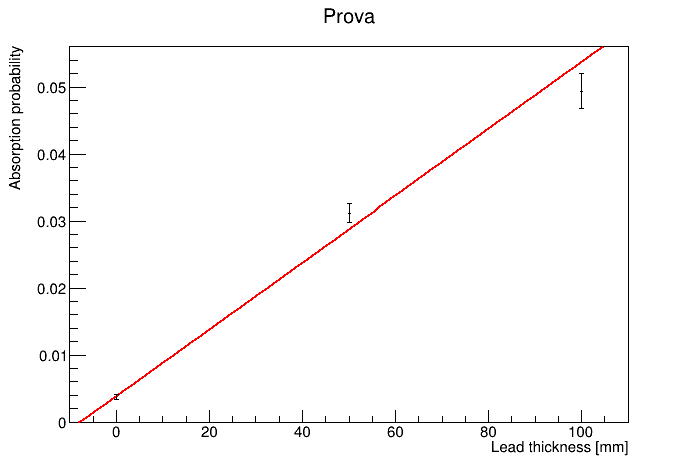}
    \caption{Absorption probability of a muon expressed as a function of the lead thickness for Rays with $\tan \alpha_{x, y} \in [-0.1, 0.1]$ (close to the vertical direction). The three points with the error bars correspond to the calibration measurements and the red line is their linear fit. This linear regression is used to convert the absorption probability into an opacity measurement for the aforementioned Rays. The absorption probability is greater than zero also for a null lead thickness because of the muon absorption by the aluminum supporting plane.}
    \label{fig:calibrazione}
\end{figure}
In figure~\ref{fig:calibrazione} the calibration measurements for all the Rays with $\tan \alpha_{x, y} \in [-0.1, 0.1]$ and their linear fit is reported.

\section{Results}

\begin{figure}
    \centering
    \includegraphics[ width=0.47\columnwidth,trim=0 0 0 40,clip] {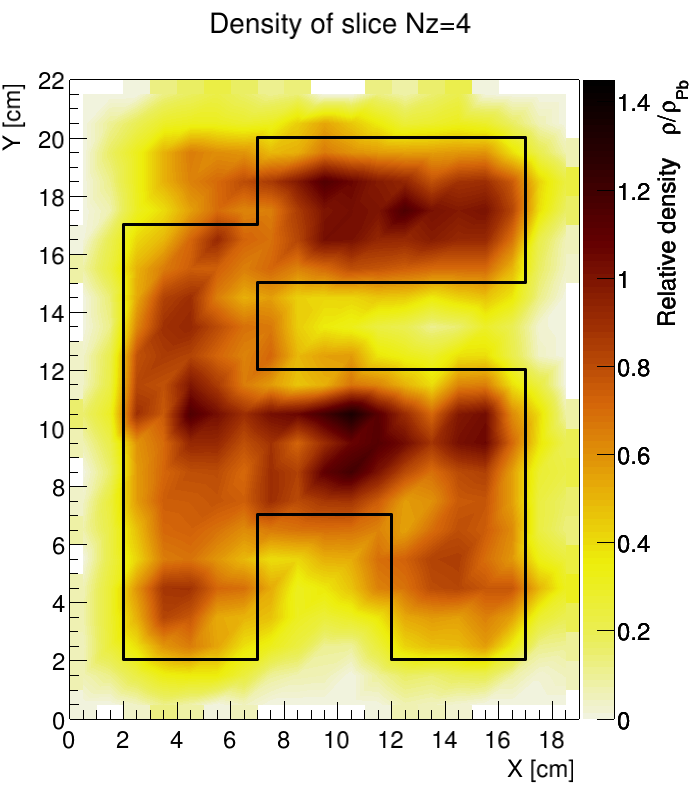}~~~
    \includegraphics[ width=0.47\columnwidth,trim=0 0 0 40,clip] {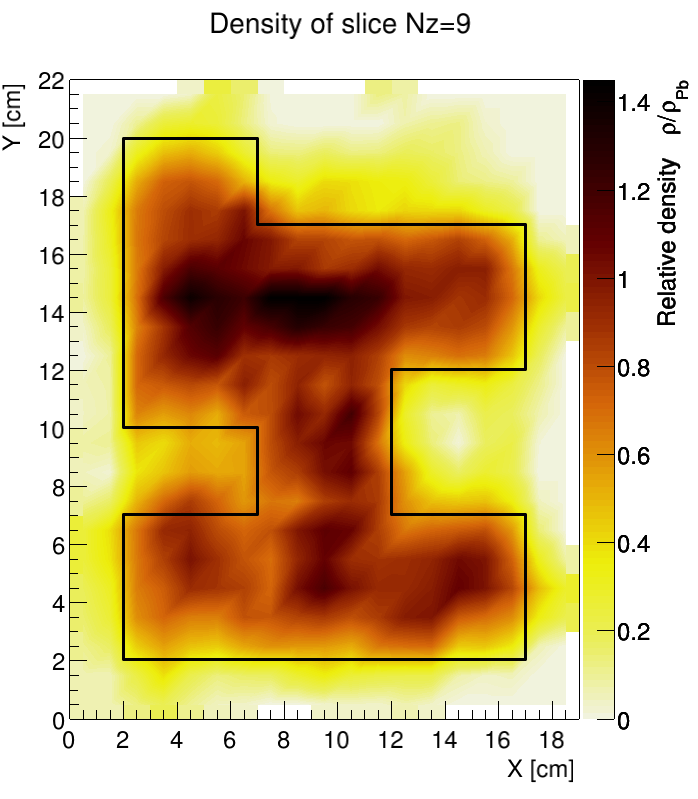}
    \caption{Slicing of the tomographic reconstruction after 10 iterations with the SART algorithm for the World1 configuration. The two slices correspond to the median planes of the two layers the target is made of. The black line denotes the real lead boundaries of the two layers (as reported in figure~\ref{fig:box_layers}). The colour scale shows the relative density compared to that of lead.}
    \label{fig:tomography1}
\end{figure}

After 10 SART iterations applied to the World1 configuration ($J=5852,\, I=3398$) %(for a computational time of 0.4 seconds)
a good agreement was obtained between the distribution of reconstructed density (see figure~\ref{fig:tomography1}) and the real one: the shape of the tomographic reconstruction reflects that of the target (see figure~\ref{fig:box_layers} for comparison) and also the density values of the denser regions are compatible with that of lead.

However, by increasing the number of iterations, the tomography loses continuity, the shape of the target is no longer recognizable and some Voxels with densities much greater than that of lead appear. The divergence of the algorithm can be explained by noting that there is a number of free parameters greater than the number of measures ($J>I$). Under these conditions it would not even be possible to use a minimization algorithm since the number of degrees of freedom ($I-J$) would be negative. Nevertheless, the SART algorithm still allows to obtain a solution to the inversion problem, that anyway diverges as the number of iterations increases.

\begin{figure}
    \centering
    \includegraphics[ width=0.47\columnwidth,trim=0 0 0 40,clip]{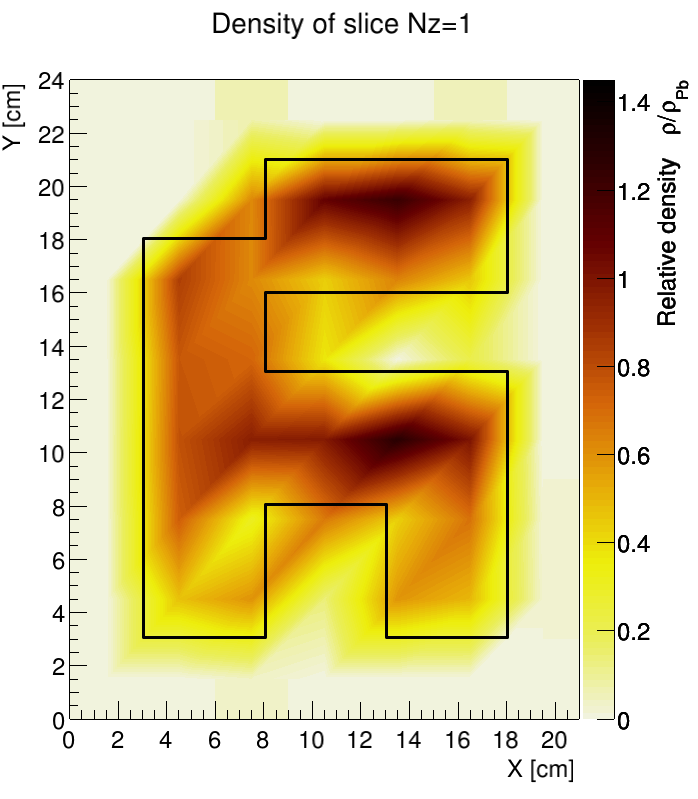}~~~
    \includegraphics[ width=0.47\columnwidth,trim=0 0 0 40,clip]{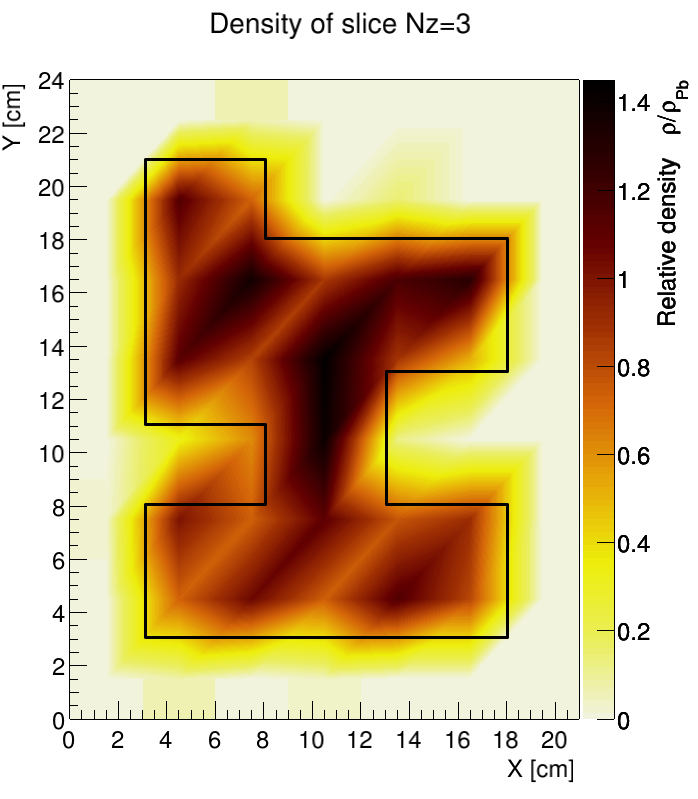}
    \caption{Slicing of the tomographic reconstruction after the Chi-squared minimization for the World3 configuration. The two slices are positioned at the median planes of the two target layers. The black line denotes the real lead boundaries (see figure~\ref{fig:box_layers} for comparison). The vertical scale indicates the slice density in units of the lead density.}
    \label{fig:tomography2}
\end{figure}

This divergence of the SART algorithm is removed by decreasing the number of unknowns and consequently increasing the size of the Voxels like in the World3 configuration (${J=336,\, I=4095}$). In this case, a Chi-squared minimization method was also used, obtaining similar results (see figure~\ref{fig:tomography2}), but with a convergence time 750 times greater than that required to perform 10 SART iterations.
%5 minutes

\section{Discussions and conclusions}
The three-dimensional distribution of density of the target was correctly reconstructed combining multiple absorption muon radiography measurements: thanks to the use of tomographic algorithms (SART and Chi-squared minimum) it was possible to reconstruct the shape and the density of the target. For this specific application, the SART algorithm has proved to be more solid being able to lead to a solution even in the case of a ill-posed problem (as in the World1 configuration).  In the case of the World3 configuration, both methods proved to be effective and no regularization was necessary for the Chi-squared minimization. This was made possible thanks to the reduction in the number of unknowns (which is an order of magnitude smaller than the number of measurements) due to the increase in the size of the Voxels, and thanks to the possibility of observing the target from multiple directions, even close to the horizontal.

In the case of muon transmission radiography, the position, direction and number of measurements are very often limited by the logistics of the installation site. For this reason it is normal to resort to regularization terms which help to make the inversion problem soluble. For the future it is therefore planned to apply these tomographic reconstruction techniques (suitably modified for the transmission technique) in cases where a series of multiple muographic measurements are available, both in the archaeological and mining fields.

\section*{Acknowledgement}
My thanks go to Dr. Sébastien Procureur of the Irfu Institute of the CEA center in Saclay who supervised the work presented in this article and revised the text. I thank my PhD tutors Prof. Raffaello D'Alessandro and Dr. Lorenzo Bonechi of the Florence section of the INFN for revising the article.

\printbibliography

\end{document}